\documentclass[conference]{IEEEtran}
\usepackage{times,amsmath,epsfig,latexsym,amssymb,psfrag}
\usepackage{paralist}
\usepackage{booktabs}
\usepackage{subcaption}
\usepackage{color}
\usepackage{setspace}
\usepackage[ruled,vlined]{algorithm2e}

\newcommand{\qed}{\nobreak \ifvmode \relax \else
\ifdim\lastskip<1.5em \hskip-\lastskip
\hskip1.5em plus0em minus0.5em \fi \nobreak
\vrule height0.75em width0.5em depth0.25em\fi}

\begin{document}

\title{Grant-Free Radio Access for Short-Packet Communications over 5G Networks }
\author{ Amin Azari$^{1,2}$, Petar Popovski$^2$, Guowang Miao$^{1}$, \v Cedomir Stefanovi\' c$^2$\\
$^1$KTH Royal Institute of Technology; $^2$Aalborg University\\
Email:\{aazari,guowang\}@kth.se, \{cs,petarp\}@es.aau.dk}

\maketitle



\begin{abstract}
Radio access management plays a vital role in delay and energy consumption of connected devices.
The radio access in existing cellular networks is unable to efficiently support massive connectivity, due to its signaling overhead.
In this paper, we investigate an asynchronous grant-free narrowband data transmission protocol that aims to provide low energy consumption and delay, by relaxing the synchronization/reservation requirement at the cost of sending several packet copies at the transmitter side and more complex signal processing at the receiver side.
Specifically, the timing and frequency offsets, as well as sending of multiple replicas of the same packet, are exploited as form of diversities at the receiver-side to trigger successive interference cancellation.
The proposed scheme is investigated by deriving closed-form expressions for key performance indicators, including reliability and battery-lifetime. The performance evaluation indicates that the scheme can be tuned to realize long battery lifetime radio access for low-complexity devices.
The obtained results indicate existence of traffic load regions, where synchronous access outperforms asynchronous access and vice versa.
\end{abstract}


\IEEEpeerreviewmaketitle

\section{Introduction}

Internet of things (IoT) 
 is expected to be integrated in cellular networks by 2020 \cite{ref2020}. 
The characteristics of IoT include: extremely high density of nodes, short payload size, and vastly diverse quality-of-services (QoS) requirements. 
Moreover, devices in most of IoT applications are battery driven, necessitating long battery lifetime \cite{e2mac}.
Thus, in contrast to the existing cellular traffic, the IoT traffic requires support for (i) massive concurrent access, (ii) high energy efficiency, and (iii) low latency with ultra-high reliability.  
The continuing growth of IoT market has encouraged mobile network operators (MNOs) to investigate evolutionary and revolutionary radio access technologies for addressing these problems \cite{revol}. 

\subsection{Literature Study}

Evolutionary schemes aim at enhancing access procedures of existing LTE networks \cite{sched}.
In existing LTE networks, devices contend over random access channel (RACH) to reserve radio resources, and then send data over granted resources.
As in most IoT application the actual data to be transmitted is in order of bits, this connectivity procedure results in unnecessary energy consumption in overhead signaling and idle listening to the base station (BS) \cite{e2mac}.
As a result, battery lifetime of connected devices will be much less than the 5G requirements, i.e. more than  10 years of battery lifetime \cite{nok}.
Also, radio access congestion is possible due to the massive number of potential connections that should be sustained concurrently \cite{cong}.
Capacity limits of RACH when serving IoT traffic and a survey of improved solutions can be found in \cite{laya}, where among the proposed solutions, access class barring and capillary networking  have been adopted in standardization \cite{acb}.

On the other hand, revolutionary solutions aim at fundamental revision of the cellular access procedures.
The development of LTE for low-cost massive IoT has been initiated in release 12 and has been continued in release 13 with introduction of narrow-band cellular IoT (NB-CIoT) \cite{ciot}.
In NB-CIoT, the bandwidth for communications and data rates has been decreased significantly in order to improve the link budget, and hence, reduce the required energy for data transmission.
However, it still suffers from required overhead signaling for synchronization, listening for ACK per messages, etc. 
A potential solution to tackle this problem is to enable grant-free communications for short-packets.
Among proposed grant-free schemes, asynchronous ALOHA has the advantage of reduced required complexity at the transmitter side 
\cite{TruAs,twod}.
To further improve the performance 
each device may replicate its packets several times,
which is exploited by the receiver through (i) decoding of packets by combining their (partially) interference-free replicas and (ii) removal of replicas of decoded packet through interference cancellation, enabling potential decoding of new packets.
Such successive interference cancellation (SIC)-based receivers for asynchronous ALOHA systems have been investigated in \cite{TruAs,ecra}.
Specifically, the solution in \cite{ecra} exploits timing offsets and replica ``diversity'', 
but the proposed receiver requires complete knowledge of the replicas position of the undetected users.
The approach in \cite{ecra} uses correlation for replica detection or robust encoding of the information of the placement of the other replicas, which is embedded in the packet header.
However, the performance of the proposed solutions decreases as the traffic load, and, thus, the amount of interference, increases.
Furthermore, the correlation in search of replicas significantly increases complexity of receiver (as discussed in section \ref{trcd}), and hence, increases the required time to detect and decode the packets; this is not consistent with the goal of reducing the experienced  delay through grant-free access. 

Another important aspect to be taken account when designing grant-free schemes is that a big portion of IoT devices are expected to be low-complexity devices with cheap oscillators.
This inevitably implies carrier frequency offset (CFO) \cite{twod}, which can potentially severely degrade the performance. In \cite{ppz}, the CFO and time offset of devices have been used for simulaneous detection of multi RFID-tags. 
In this paper, we propose to exploit CFO as another source of diversity and develop a SIC-enabled time and frequency asynchronous ALOHA-based grant-free access, which can support multitude of low-complexity IoT devices.

\subsection{Contributions}

The contributions of the paper are the following:
\begin{itemize}
\item
Development of a SIC-enabled time/frequency asynchronous radio access scheme for grant-free communications. Development of a  collision resolution scheme utilizing  time/frequency domain asynchronism.

\item
Development of a closed-from statistics of two-dimensional (i.e., time-frequency) interference, and derive expressions for outage probability, expected battery lifetime, experienced delay, spectral efficiency, and energy efficiency of the network. 
\item
Evaluation of fundamental tradeoffs for access protocols with short packets. 
\item
Identification of operating regions in terms of traffic load in which asynchronous access outperforms synchronous and granted access.
\end{itemize}

The remainder of the text is structured as follows.
The system model is described in the next section.
In Section~III, the proposed transceiver design is presented.
Performance indicators are modeled analytically in Section~IV, and performance tradeoffs are investigated in Section~V.
Simulation results are presented in Section~VI.
The concluding remarks are given in Section~VII.

\section{System Model}

\begin{figure}[t!]
        \centering
 \begin{subfigure}[t]{0.5\textwidth}
        \centering
  \includegraphics[width=3.5in]{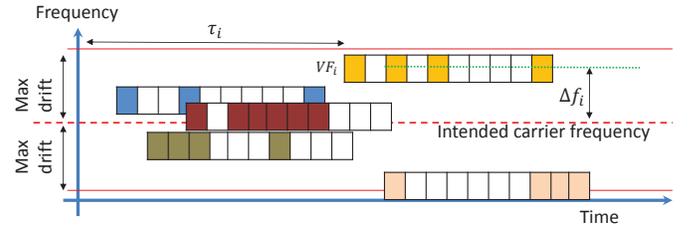}
\caption{Received virtual frames at the BS. The colored slots contain packet replicas.} \label{pr}
\end{subfigure}\\        
 \begin{subfigure}[t]{0.5\textwidth}
        \centering
  \includegraphics[width=3in]{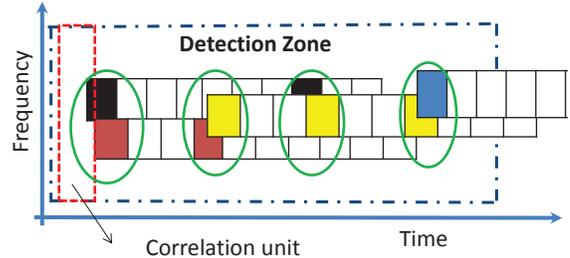}
\caption{A detection zone with 4 unsolvable collisions} \label{dz}
\end{subfigure}\\        

  \caption{Time-frequency asynchronous ALOHA}\label{figAA}
\end{figure}

We consider a single cell serving multitude of IoT devices.
Upon having a packet to transmit, the $i$-th device assumes a virtual frame (VF) consisting of $M$ slots, each with duration {$T_p^i$}, where $T_p^i$ is the time-duration of a packet transmission of device $i$. Then, packet is sent immediately at the first slot of the VF, and the $N_i-1$ replicas of the packet are sent in $N_i-1$ randomly selected slots out of $M-1$ remaining slots, as depicted in Fig.~\ref{pr}; where $N_i$ is randomly chosen from $\{1,\cdots,M\}$. A quasi-static fading channel model is assumed, which means channel gain is constant over a VF. The transmitted packet is intended to be modulated over the carrier frequency (CF), denoted by $f$, which is the same for all devices.
As low-complexity sensors with cheap oscillators will be an essential part of future networks, CFO will be inevitable.
Indeed, it is expected that in ultra-narrowband (UNB) systems, the level of CFO is expected to be several orders higher than the communications bandwidth \cite[section~3.2.2]{thes}. 
Denote the actual carrier frequency that the $i$th transmitter uses for data transmission, and its drift from the intended carrier frequency as $f_i$ and $\Delta f_i = f_i - f$.
While frequency drifts in different wakeup epochs of operation of devices are expected to be different, $\Delta f_i$  is, in essence, constant during one virtual frame, i.e. for $M T_p^i$ seconds \cite{cfo}.

The same transmission strategy is uncoordinatedly used by all devices with pending data transmission, where timing offsets, CFOs and number of transmitted replicas are independent among devices.
Thus, overlapping of packet replicas sent by devices' is inevitable, as depicted in Fig.~\ref{pr}.


\section{Transceiver Design}\label{trcd}

Fig. \ref{pr} represents packet reception at the receiver. 
 As the timings and CFO of virtual frames are unknown at the receiver side, a sliding detection zone is used, see Fig.~\ref{dz}.
Following the design in \cite{TruAs,ecra}, we assume that the time duration of the detection zone is a factor of VF's length,
When the traffic load (i.e., the overall number of replicas) in the detection zone is low, the receiver can simply decode replicas that are not in collision, perform interference cancellation, and repeat the procedure for the new ``uncovered'' replicas of other packets.
However, as the load increases, it may happen that all packet replicas of all devices are in collision.
In this case, correlation has been proposed to find the position of replicas \cite{ecra}.
This approach, which we refer to as blind correlation, significantly increases the complexity, as follows.

Consider the scenario in Fig.~\ref{dz}, where all packet replicas of all 4 devices are in collision.
Using blind correlation, receiver needs to select a correlation ``unit'', depicted in Fig. \ref{dz} by a red-colored rectangle, slide and correlate this unit with all taken samples inside the detection zone.
In the next step, receiver moves the correlation unit in time-frequency, then correlates this unit with all samples in the detection zone; this procedure is repeated for all possible correlation units.
After doing all these correlations, there would be no results in case of Fig. \ref{dz}, because the colliding replicas are different in each collision event in Fig.~\ref{dz}, as well as the respective CFOs, are different.
Even the position of replicas are found, the receiver only use equal gain combining, which is inefficient due to the different level of interference that each of them is suffering \cite{mrcegc}.
As a result, the potential for resolution of transmitted packets is low. 

\subsection{The Proposed Transceiver Design}

\begin{figure}[t!]
 \centering
 \includegraphics[width=3.7in]{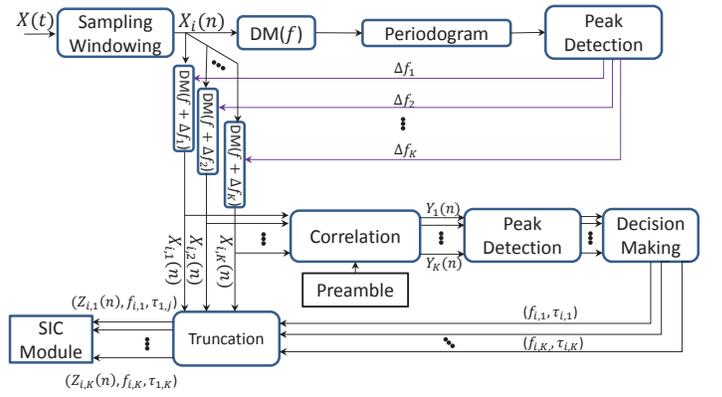}
\caption{The proposed receiver design. $i$ is the event index. DM($f$) represents demodulation with frequency $f$.} \label{op}
\end{figure}

Summarizing the above discussions, we need to design fast, yet accurate collision resolution procedure.
Towards this end, we add a known preamble of length $N_z$ to each transmitted packet.
The preamble can be selected from Zadoff-Chu sequences, which have very good autocorrelation properties.
At the receiver end, we sample the arriving signal at rate $F_s$, searching for the (potentially collided) signals, see Fig.~\ref{op}. The choice of $F_s$ introduces a tradeoff to the system performance, because a sampling rate higher than the Nyquist rate increases both receiver's cost and collision resolution capability. Once the presence of the signal is detected, to which we refer to as an event, the receiver jointly processes samples organized in a time frame with length of $T_f$, where $T_f<\min⁡\{T_{\max},EoE\}$.
$T_{\max}$ is a design parameter related to the tolerable delay in data processing, while EoE denotes the end of event, i.e., when the presence of the signal can not be detected any more and the channel is sensed to be idle again.
For example, in Fig.~\ref{dz} there are 4 such events, each consisting of 2 collided transmissions.

The samples in the time frame are processed using a periodogram module, which aims at finding periodic components in the signal, and returns found carrier frequencies.
Denote the number of found carrier frequencies (which are determined by CFOs of the contending devices) as $K$ (see in Fig. \ref{op}).
Then, samples from the time frame are demodulated and correlated with the preamble $K$ times (the preamble is known at the receiver).
Each of the correlations returns some peaks.
Consider $Y_j(n)$, the  output of  correlation of $X_{i,j}$, which is the demodulated version of $X_i(n)$  by $f+\Delta f_i$, with the preamble.
In $Y_j (n)$, the respective peak of $j$-th CFO has been located at the right timing offset, while the respective peaks of other CFOs have been shifted.
The reasons behind these shifts are further discussed in Section~\ref{shif}; here we note that the level of shifted peaks can be as high as the original peak, or even higher, if the length of preamble sequence is short, which may be the case in IoT applications with short-packet lengths.
The task of the peak detection  module in Fig.~\ref{op} is to report the set of detected peaks. The task of decision making   module is to detect and remove shifts of time-offsets of already detected peaks, to be discussed in detail in Section~\ref{shif}, and to report the set of $K$ found time offsets (respective to the $K$ found CFOs). 
Then, the respective demodulated sequence of each carrier frequency, e.g. $X_{i,j}(n)$ for $f_i$,  is truncated from $\tau_{i,j}$ to the length of a packet and is fed to the SIC module along with its carrier frequency and time offset, i.e. $(Z_{i,j}(n), f_{i,j}, \tau_{i,j} )$ are fed to the SIC module.

\subsection{The Proposed SIC Module}\label{sic}

The SIC module continuously receives and saves demodulated sequences related to processed events, and their respective carrier frequencies and time offsets, i.e. the set of $(Z_{i,j}(n),f_{i,j}, \tau_{i,j})$. 
Then, it tries to decode each sequence.
If the sequence,  i.e. the supposedly contained packet replica, is decoded successfully, the location of the other replicas becomes known, and hence, these are removed.
If the packet replica cannot be decoded correctly, SIC module tries to find its replicas by search in other processed and stored events containing the same carrier CFO.
If the other replicas are also in collision, the SIC module can combine them.
Thanks to the derived set of of CFOs and time offsets, we have the time/frequency map of collisions, and hence, it is possible to figure out the level of interference that each replica is suffering from, and hence, we can use selection combining (SC) or maximum ratio combining (MRC) in order to improve the performance.
The former consists of merging successfully received parts of replicas together to construct the original packet.
The latter consists of combining whole replicas, taking into account the level of interference in each replica, as explained Section~\ref{pmt}. 
After combining, SIC module again tries to decode the combined packet.
If decoding succeeds, the receiver removes all replicas of the decoded packet, which lowers the level of interference in the other demodulated sequences (i.e., processed events) and provides for easier decoding of other packets.
If decoding fails, the demodulated sequence is stored for further processing, and receiver  slides the detection zone and tries to decode newly arrived events. In case that the subsequent decodings lower the level of interference in the previously stored collisions, new decoding attempts will be made.

\subsection{Processing of Detected Peaks}\label{shif}

\begin{figure}[t!]
 \centering
 \includegraphics[width=3in]{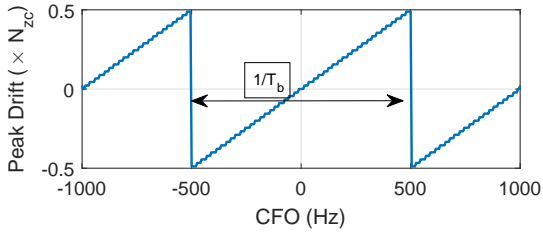}
\caption{Peak drift with CFO ($N_{zc}=45, Tb=1\, \text{ms}$). } \label{pd}
\end{figure}

We first elaborate on the reason behind having side peaks when we correlate a preamble with a sequence that contains the same preamble with CFO.
If we take cross correlation of a preamble sequence, i.e. $P(n), n\in\{0,\cdots,N_{zc}-1\}$, with itself, the result will be a sequence of length $2N_{zc}-1$, i.e. $m\in\{1,\cdots,2N_{zc}-1\}$, with a peak at $N_{zc}+1$.
Denote by $\tilde P(n)$ a modulated version of $P(n)$ with carrier frequency $\Delta f_i$, i.e. $\tilde P(n)=P(n)e^{j2\pi \Delta f_i n T_b}$, $\forall n\in\{0,N_{zc}-1\}$, where $T_b$ is the bit duration.
Taking cross correlation of $P(n)$ with $\tilde P(n)$, one sees the peak location changes periodically between $-\lfloor N_{zc}/2\rfloor$ and $\lfloor N_{zc}/2\rfloor$, as discussed in \cite{shif}.
Given $T_b$ and $N_{zc}$ as characteristics of the system, position of shifted peak can be derived as a function $\Delta f_i$, denoted by $Q(\Delta f_i)$, as depicted in Fig.~\ref{pd} for $N_{zc}=45, T_b=1 ms$.
This function can be evaluated once and stored in a lookup table for a given CFO range, to be used in the decision making module.

The decision making module decides which subset of detected peaks represents time offsets of the replicas.
In this module, a successive peak cancellation (SPC) function is used.
Denote  the set of received peaks from peaks detection module as $\{\mathcal T_{1},\cdots,\mathcal T_{K}\}$, where $\mathcal T_j$ is the set of detected peaks in $Y_{j}(n)$, and $Y_{j}(n)$ is the result of correlation of $X_{j}(n)$ with the preamble, as in Fig.~\ref{op}.
The SPC function searches over $\mathcal T_{j}:$s and makes a map of peaks and their shifted positions.
For example, given $p_j$ as  a candidate peak position\footnote{$p_j$ represents time offset of a peak w.r.t. the reference time in processing of the respective event.} in $\mathcal T_j$, SPC checks $\mathcal T_{j+k}$ to see if it contains a peak at $Q\big(-(\Delta f_{j+k}-\Delta f_{j})\big)$ ,$\forall k\in\{1,\cdots,K\}\setminus j$, corresponding to the repeated occurrence of $p_j$.
If $p_j$'s  repetitions can be found in $\mathcal T_k$s, then $p_j$ is validated, else it is removed from $\mathcal T_j$.

\section{Performance metrics and tradeoffs}\label{pmt}

\subsection{Reliability Analysis}\label{ou}

Here, we formulate the success probability of packet transmissions, as a function of system and traffic parameters.
To make the analysis tractable, radio channel is modeled by a distance-dependent variable, and it is assumed that devices use channel-inversion transmit-power control to achieve a constant signal-to-noise ratio at the receiver.
The packet transmission duration is $T_p=D/\big[W\log_2 (1+\gamma/\Gamma)\big]$, where $D$ is the packet length, $W$ is the bandwidth, $\gamma$ is the required SNR at the receiver, and $\Gamma$ is the SNR gap between channel capacity and a practical coding and modulation scheme. Furthermore, the number of transmitted replicas per  packet is assumed to be $N$, for all devices.
As in \cite{twod}, we assume that transmitted energy is uniformly distributed over its time-frequency support, i.e. over a rectangle of size $W\times T_p$.
Then, the ratio between energy of contained in a replica and total energy of its interference and noise is modeled as:
\begin{align}
\text{SINR}&=\frac{\rho W T_p }{\sum_{k} \rho S_k+N_0 W T_p}=\frac{1}{\frac{1}{W T_p}\mathcal M+\gamma^{-1},}
\end{align}
where $\rho$ is the transmitted energy density over the time-frequency support, $N_0$ is the energy density of noise, $\mathcal M=\sum_k S_k$,  and  $S_k$ is the area of the ``overlap'' between replica of $k$-th interfering packet and the replica of the original packet.
Thus, the problem reduces to finding the set of interfering replicas of other packets and $S_k$.
Denote by $F_m$ the maximum drift from the carrier frequency $f$, and assume that CFO is uniformly distributed in $[-F_m,F_m]$, i.e. the available frequency spectrum is $[f-F_m+W/2,f+F_m+W/2]$.
Then, for a replica which starts at $t_0$ with frequency offset of $0$, and hence, spans over $(t_0,t_0+T_p)\times (f-W/2,f+W/2)$, 
the vulnerable zone is $(T_p-t_0,t_0+T_p)\times (f-W,f+W)$. This means that any packet transmission starts in $t_x$ with carrier frequency $f_x$, where  $t_x\in (T_p-t_0,t_0+T_p)$ and  $f_x\in (f-W,f+W)$,  interferes with the intended packet.

Assuming that the number of interfering replicas of other packets in the vulnerable period is $n$, the conditional cumulative distribution function (CDF) of $\mathcal M$, i.e. $F_M(n)(x)$, is:
\begin{align}
\Pr (S_{k} >s ) = & 4\int\nolimits_{0}^{T_p-\frac{s}{W}}\bigg[\int\nolimits_{0}^{W-\frac{s}{T-u}}\frac{1}{2T_p}\frac{1}{2F_m} dv\bigg] du \\
 = & \frac{4}{4T_p F_m}\int\nolimits_{0}^{T-\frac{x}{w}}[W-\frac{s}{T-u}]du \\
= & \frac{1}{T_p F_m}\bigg[W(T -\frac{s}{W})+s\ln(\frac{s}{TW})\bigg], \\
 F_{S_k}(s)  = & 1-\frac{1}{T_p F_m}\bigg[W(T-\frac{s}{W})+s\ln(\frac{s}{TW})\bigg],\\
F_{ \mathcal M(n)}(s) = & F_{S_{1}}(s)*f_{S_{2}}(s)\cdots*f_{S_{n}}(s),\label{cdf}
\end{align}
where $F_X(x) = \Pr (X\le x)$.

We proceed by deriving the unconditional CDF of $\mathcal M$.
Denote the aggregated packet transmission rate of devices as $g$, where  $g\approx N\lambda/(1-\mathcal P_o)$ \cite[section~21.1.2]{gb}, $\lambda$ is the aggregated new packet arrival rate at devices, and $\mathcal P_o$ is the probability that a packet cannot be decoded in its VF. 
The unconditional CDF is:
\begin{align}
&F_{\mathcal M}(s)=\sum\nolimits_{i=1}^{\infty} F_{ \mathcal M(i-1)}(s){[g2T_p]^i e^{-g2T_p}}/{i!}.
\end{align}
Taking into account that the expected number of interfering replicas in the vulnerable period is $\bar n=\lceil g2T_p\rceil$, one may simplify the analysis by substituting  $n$ with $\bar n-1$ in \eqref{cdf}  to derive the unconditional CDF.
Further, the probability of outage is derived as:
\begin{align}
\mathcal P_o& = \Pr (\text{SINR}< S_t)= \Pr ([{ \mathcal M/{W T_p}+\gamma}]^{-1}< S_t),\label{pou}\\
&= \Pr (\mathcal M> WT_p\big[1/S_t\text{-}1/\gamma\big]=1-F_{\mathcal M}(WT_p\big[1/S_t\text{-}\gamma\big]),\nonumber
\end{align}
where $S_t$ is the threshold SINR for correct decoding. 
In case every packet is transmitted with the same number of replicas $N$, we use the minimum mean-square error (MMSE) criterion, and combine replicas based on the level of interference that they suffer from.
Denote by $\mathcal M_i$ the sum of intersection areas of interfering packets with the $i$-th replica, $i\in\{1,\cdots,N\}$. Then, we have:
\[\left[ {{Y_1}, \cdots ,{Y_N}} \right]^T = \left[ {X, \cdots ,X} \right]^T + \left[ {{\Omega _1}, \cdots ,{\Omega _N}} \right]^T,\]
in which ${\bf a}^T$ represents transpose of vector $\bf a$, and  $X$, $Y_i$ and $\Omega_i$ represent the intended  {signal}, observation, and noise plus interference, respectively.
The powers of $\Omega_i$ and the intended signal are denoted by $\sigma_i = N_0WT_p+\mathcal M_iWT_p\rho$ and $\sigma_x=\rho WT_p$, respectively.
The optimal combining weight coefficients by MMSE criterion are:
\[\left[ \begin{array}{l}
{w_1}\\
 \vdots \\
{w_N}
\end{array} \right] = {\left[ {\begin{array}{*{20}{c}}
{\sigma _x^2 + \sigma _1^2}& \cdots &{\sigma _x^2}\\
 \vdots & \ddots & \vdots \\
{\sigma _x^2}& \cdots &{\sigma _x^2 + \sigma _N^2}
\end{array}} \right]^{ - 1}}\left[ \begin{array}{l}
\sigma _x^2\\
 \vdots \\
\sigma _x^2
\end{array} \right],\]
and hence, the combination to be decoded is: 
$Y_c=[w_1,\cdots,w_N][Y_1,\cdots,Y_N]^T$.
The resulting SINR is then the sum of SINR of packets, and  {the probability of outage} is:
\begin{align}
\mathcal P_o&= \Pr (N \text{SINR}< S_t)= \Pr (\frac{N}{\mathcal M/{W T_p}+1/\gamma}< S_t),\\
&= \Pr(\mathcal M> WT_p\big[N/S_t\text{-}1/\gamma\big]=1\text{-}F_{\mathcal M}(WT_p\big[N/S_t\text{-}1/\gamma\big]).\nonumber
\end{align}
In case without replica combining, when the decoding is attempted for each replica individually, the probability of outage is:
\begin{align}
\mathcal P_o&=\prod\limits_{i=1}^{N} \Pr ( \text{SINR}< S_t)=\big[ \Pr (\frac{1}{ \mathcal M\big/{W T_p}+1/\gamma}< S_t)\big]^N,\nonumber\\
&=\big[1\text{-}F_{\mathcal M}(WT_p\big[N/S_t-1/\gamma\big])\big]^N.\nonumber
\end{align}

\subsection{Delay Analysis}

The average experienced delay from packet arrival at a device to successful reception at the BS is:
\begin{align}
\text{ED}=\sum\nolimits_{i=1}^{\infty} [M T_p+T_{ack}]P_o^{i-1}[1-P_o]-T_{ack},
\end{align}
in which we have assumed that a device retransmits the packet if it doesn't receive ACK within $T_{ack}$ seconds.

\subsection{Battery Lifetime}

For most reporting applications, the packet generation process at each device can be modeled as a Poisson process, and hence, energy consumption of each device can be seen as a semi-regenerative process where the regeneration points are located at the end of each successful data transmission epoch.
Denote the battery capacity of the $i$th device at the reference time as $E_0$, the average time between two data transmissions as $T_r$, and the average packet size as $D$.
Also, power consumption of node $i$ in the listening and transmission modes are denoted as $P_c$ and $\alpha \tilde P_t+P_c$  respectively, where $P_c$ is the circuit power consumed by electronic circuits, and $\alpha$ is the inverse power amplifier (PA) efficiency.
As the required SNR at the BS is $\gamma$, the transmit power of device $i$ located at distance $r_i$ from the BS is modeled as:
\begin{equation}\label{powe}
P_{t_i}={\gamma N_0 W \Gamma }r_i^{\sigma}/\mathcal G ,
\end{equation}
where $\mathcal G$ is the multiplication of transmit and receive antenna gains, $\sigma$ is the path loss exponent, $\Gamma$  the SNR gap between channel capacity and a practical coding and modulation scheme.
Assuming the uniform distribution of devices in the cell, the PDF of the distance between a device and the BS is $f(r)=\frac{2r}{R_c^2}$, where $R_c$ is the cell radius and $r$ is the communications distance.
The long-term average of the required transmit power is then:
\begin{align}
\bar P_{t}=&\int\nolimits_{0}^{R_c}\frac{\gamma N_0 W \Gamma r^{\sigma}}{\mathcal G}\frac{2r}{R_c^2}dr=\frac{2R_c^{\sigma}\gamma N_0 W \Gamma }{\mathcal G[\sigma+2]}.\label{pta}
\end{align}
Now, we define the expected {\textit { battery lifetime}} at the regeneration point as the product of reporting period and the ratio between remaining energy and the average energy consumption per reporting period, as follows:
\begin{equation} \mathcal L =\frac{E_{0}{ T_r}}{E_{st}+\frac{1}{P_o}\big[[P_c+\alpha \bar P_t] NT_p+P_c (M-N)T_p+P_c T_{ack}\big]} \label{lif} ,\end{equation}
where $T_{ack}$ is the average waiting time for receiving ACK, $N$ is the number of replicas transmitted per packet, $M$ number of slots in a VF, and  $E_{st}$ the average static energy consumption in each reporting period for data processing etc. 

\subsection{Energy Efficiency}

The energy efficiency of devices in uplink communications  in terms of Bit/Joule can be approximated as the ratio between number of useful transmitted bits in $MT_p$ seconds to the consumed energy in that interval, as follows:
\begin{align}
\text{EE}=&\frac{ {\lambda MT_p}  [D-D_{oh}]}{ \frac{\lambda MT_p}{1-P_o} \big[[P_c+\alpha \bar P_t] NT_p\text{+}P_c [M-N]T_p\text{+}P_c T_{ack}\big]},\nonumber\\
=&\frac{(1-P_o)[D-D_{oh}] }{\big[[P_c\text{+}\alpha \bar P_t] NT_p+P_c [M-N]T_p\text{+}P_c T_{ack}\big]},\nonumber
\end{align}
in which $D_{oh}$ denotes number of overhead bits in a packet, e.g. for synchronization and cyclic prefix.

\subsection{Spectral Efficiency}

The spectral efficiency of network in terms of Bit/Sec/Hz can be approximated as the ratio between number of  {successfully received bits} in $MT_p$ seconds versus the time-frequency reserved radio resources in that interval, as follows:
\begin{align}
\text{SE}=\frac{{\lambda MT_p}[D-D_{oh}] (\text{Bit})}{2MT_p[F_m+W/2] \text{Sec.Hz}}=\frac{\lambda[D-D_{oh}] }{2[F_m+W/2]} \text{Bit/S/Hz}.\nonumber
\end{align}

\section{Tradeoffs in Radio Access Design for Massive Short-Packet Communications}\label{trf}

{From an overall system perspective, we aim at minimizing the costs of the access network, maximizing spectral efficiency, maximizing the energy efficiency of communications, minimizing the experienced delay in data transmission, and prolonging battery lifetime of devices.}
These objectives cannot be treated separately because they are coupled in conflicting ways.
In the following, we highlight some of these tradeoffs. 

From the expressions derived in the previous section, and the system design in Section~\ref{trcd}, there is an obvious tradeoff between energy consumption/battery lifetime of devices and costs of the access network.
Costs of the access network include deployment (CAPEX) and operational expenses (OPEX), and reducing energy consumption of devices needs more investment in CAPEX and/or OPEX of the access network.
For example, \eqref{lif} shows that the expected battery lifetime increases by decreasing the transmit power and outage probability.
Further, \eqref{pta} shows that transmit power can be decreased by denser deployment of the BSs, and thus, shortening device-BS distances, which increases the CAPEX.
\eqref{pou} shows that the outage probability can be decreased by increasing the available radio resources or using a receiver that is able to perform improved decoding/combining of replicas in collisions; both increase the OPEX of the access network. 
Furthermore, as noted in Section~\ref{shif}, detecting replicas in collisions requires either long synchronization preambles or sophisticated receivers to perform processing of derived peaks from cross correlations.
The former increases the packet size, and hence, increases the collision probability, which in turn implies less energy- and spectral-efficiency, as well as shorter battery lifetime. 
The latter increases complexity of receivers, as well as the decoding delay.
Finally, increase in the available bandwidth in order to further exploit the CFO of the devices increases the access network costs. 
Further tradeoffs can be seen in tuning transmission power of replicas to achieve ultra-high reliability or ultra-long battery lifetime, which are elaborated in the next section.

\begin{table}[t!]
\centering \caption{Simulation Parameters   }\label{sim3}
\begin{tabular}{p{4 cm}p{4 cm}}\\
\toprule[0.5mm]
{\it Parameters }&{\it Value}\\
\midrule[0.5mm]
Cell outer and inner radius &  $1000$, 50 m\\
Number of devices & $10000$  \\
Pathloss & $128.1+37.6\log (\frac{d}{1000})$\\
Interference+other losses & 20 dB\\
$W; F_m; F_s$& 200; 100; 4000 Hz\\
$P_c$; $P_t^{\min}$; $P_t^{\max}$& 1; 1; 100 mW\\
$T_b; T_p$&10; 500 mSec\\ 
Required SNR ($\gamma$)& 6 dB\\
$D; D_{oh}$&100; 50 bits\\
Modulation& Nonnegative 4-PAM\\
$\mathcal D_{synch}; \mathcal E_{synch}; E_s$& 2 Sec; 6 mJoule; 1 mJoule\\
$M; N_{zc}$ & $2N, \forall N>1; 23$\\   
\bottomrule[0.5mm]
\end{tabular}
\end{table}
  

 \begin{figure}[t!]
        \centering  
 \begin{subfigure}[t]{0.5\textwidth}
         \centering
                \includegraphics[trim={0.3mm 0mm 0.3mm 3mm},clip,height=1.75in,width=3.5in]{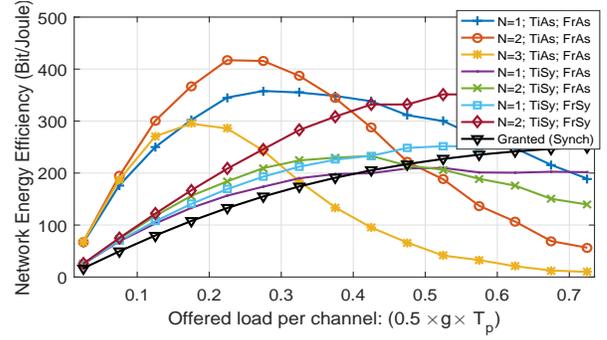}
                \caption{Network energy efficiency analysis}\label{ee}
\end{subfigure}\\    
\begin{subfigure}[t]{0.5\textwidth}
        \centering
                \includegraphics[trim={0.3mm 0mm 0.3mm 1mm},clip,height=1.75in,width=3.5in]{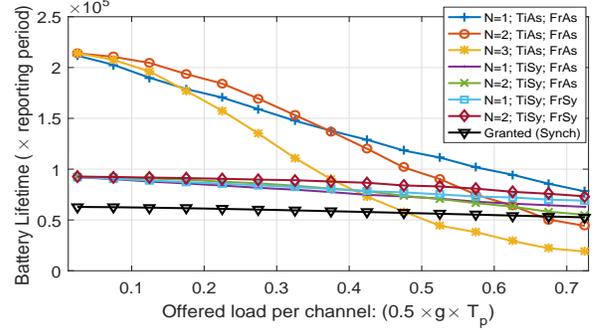}
                \caption{Average battery lifetime analysis}\label{bat}
\end{subfigure}\\   
\begin{subfigure}[t]{0.5\textwidth}
        \centering
                \includegraphics[trim={0.3mm 0mm 0.3mm 1mm},clip,height=1.75in,width=3.5in]{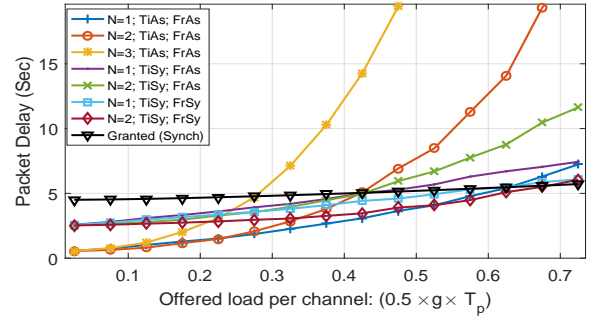}
                \caption{Packet delay analysis}\label{del}
\end{subfigure}\\   
 \begin{subfigure}[t]{0.5\textwidth}
         \centering
                \includegraphics[trim={0.3mm 0mm 0.3mm 1mm},clip,height=1.75in,width=3.5in]{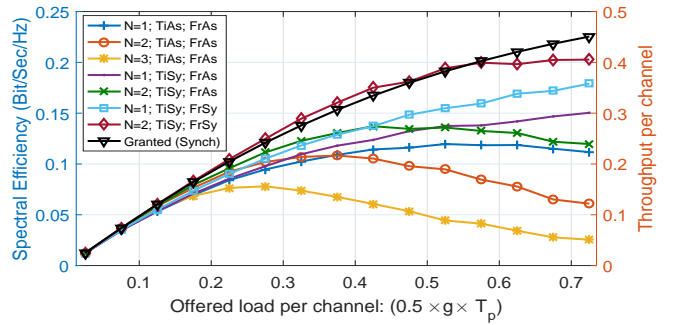}
                \caption{Throughput and spectral efficiency analysis}\label{se}
\end{subfigure}\\    
  \caption{Energy, delay, and throughput analysis.}\label{perf}
\end{figure}

\begin{figure}[!t] 
        \centering
                \includegraphics[trim={0.3mm 0mm 0.3mm 1mm},clip,height=1.75in,width=3.5in]{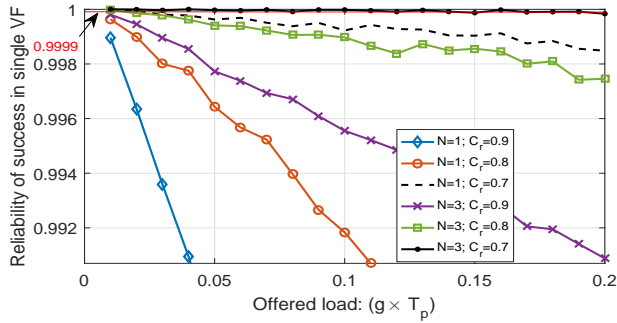}
                \caption{Reliability analysis}\label{suc}
\end{figure}

 \section{Performance Evaluations}\label{sim}
 
The system model implemented in this section is based on the uplink of a single cell with IoT traffic, with randomly distributed devices according to a spatial Poisson point process.
The  simulation parameters are listed in Table~\ref{sim3}, in which $F_s$ denotes sampling frequency, and  $\mathcal D_{synch}$ and $\mathcal E_{synch}$  represent the time and energy spent for time/frequency synchronization.

The proposed scheme can be tuned to provide extremely high energy efficiency or reliability, or a high level of both of them.
In Fig. \ref{perf}, we investigate the case ultra-high  battery lifetime is required, and hence, for $N>1$ the transmission power for each packet replica is $1/N$ of the total power that device invests in packet transmission, which is the same for all $N$.
FrAs, FrSy, TiAs, and TiSy denote frequency and time asynchronicity/synchronocity, respectively.
I.e., TiSy means that the devices are slot synchronized, while FrSy means that CFOs of the devices can take equally-spaced discrete values, i.e. the devices are sub-channel synchronized, where the channels are spaced each 200 Hz ($W=200$ Hz). Also, the black-colored curve represents the granted-access scheme in which, devices content over a random-access (RA) channel for resource reservation (10 RA resources are provided each 2 seconds), and   successful nodes transmit their packets collision-free over the data channel. The {\textit x-axis} is Fig. \ref{perf} represents the offered load per channel defined as $\frac{W}{2F_m+W}gT_p=0.5 gT_p$.

Fig.~\ref{ee} illustrates energy efficiency in uplink communications for IoT traffic versus offered load.
Obviously, in low to medium traffic load regimes, grant-free access with 2 replicas achieves the highest energy efficiency.
Fig.~\ref{bat} shows the battery lifetime performance; evidently, the battery lifetime using proposed  grant-free access has been extended by 100\% in the low to medium traffic-load regimes. 
In medium to high traffic-load regimes, number of collisions among packets transmitted using grant-free access increases, which in turn results in decreasing energy efficiency and battery lifetime.
The same fact can be seen in Fig.~\ref{del}, where packet delay using grant-free access is much lower than for  the granted access for low to medium traffic load.
Finally, Fig.~\ref{se} represents the throughput and spectral efficiency of networks versus traffic load.
It can be seen that having time and frequency synchronism increases spectral efficiency, as the collisions happen in a more controlled manner; the same insight can be traced back to pure and slotted ALOHA systems. 
Finally, the above figures also show that there regions of the traffic load in which grant-free access outperforms granted access in terms of delay and energy efficiency, and vice versa. 

Finally, Fig.~\ref{suc} represents the reliability, i.e., the probability of success of packet transmission as a function of the traffic load, for varying $N$ and the forward error correction coding rate  $C_r$.\footnote{It is assumed that the packet will be decoded if a replica combining reconstructs the fraction of its content that is up to coding rate.}
The transmission power of a replica is assumed to be the same, no matter how many replicas are sent.
Fig. \ref{suc}, shows that very high reliability, e.g. $99.99\%$ and $99.999\%$, can be guaranteed in low traffic load regions.

\section{Conclusions}

An asynchronous  grant-free radio access scheme has been proposed for low-complexity IoT devices.
The scheme aims at providing a low delay and energy consumption profile for short packet communications, by removing the synchronization/reservation requirements at the cost of sending several packet copies at the transmitter side and more complex signal processing at the receiver side. 
Closed-form expressions of key performance indicators have been derived.
It has been shown that by tuning the transmission parameters, one can achieve very long battery lifetime or highly reliable access with bounded delay for low-complexity devices.
Also, the regions of the traffic load in which synchronous/asynchronous access perform favorably have been investigated. 
The simulation results have verified the performance of the proposed system for short packet transmissions. 
Finally, we note that the proposed approach has the potential to be used in other asynchronous access solutions, e.g., in satelite communications. 

\section*{Acknowledgment} \label{acknow}

The research presented in this paper was supported in part by the Danish Council for Independent Research, grant no. DFF-4005-00281 and in part by the European Research Council (ERC Consolidator Grant Nr. 648382 WILLOW) within the Horizon 2020 Program.


%
%

\ifCLASSOPTIONcaptionsoff
  \newpage
\fi



\begin{thebibliography}{10}
\providecommand{\url}[1]{#1}
\csname url@samestyle\endcsname
\providecommand{\newblock}{\relax}
\providecommand{\bibinfo}[2]{#2}
\providecommand{\BIBentrySTDinterwordspacing}{\spaceskip=0pt\relax}
\providecommand{\BIBentryALTinterwordstretchfactor}{4}
\providecommand{\BIBentryALTinterwordspacing}{\spaceskip=\fontdimen2\font plus
\BIBentryALTinterwordstretchfactor\fontdimen3\font minus
  \fontdimen4\font\relax}
\providecommand{\BIBforeignlanguage}[2]{{%
\expandafter\ifx\csname l@#1\endcsname\relax
\typeout{** WARNING: IEEEtran.bst: No hyphenation pattern has been}%
\typeout{** loaded for the language `#1'. Using the pattern for}%
\typeout{** the default language instead.}%
\else
\language=\csname l@#1\endcsname
\fi
#2}}
\providecommand{\BIBdecl}{\relax}
\BIBdecl

\bibitem{ref2020}
M.~R. Palattella \emph{et~al.}, ``Internet of things in the {5G} era: Enablers,
  architecture, and business models,'' \emph{IEEE J. Sel. Areas Commun.},
  vol.~34, no.~3, pp. 510--527, March 2016.

\bibitem{e2mac}
G.~Miao, A.~Azari, and T.~Hwang, ``{$E^{2}$ -MAC}: Energy efficient medium
  access for massive {M2M} communications,'' \emph{IEEE Trans. Commun.},
  vol.~64, no.~11, pp. 4720--4735, Nov 2016.

\bibitem{revol}
Ericsson, Huawei, NSN, and {\it et al}., ``A choice of future {M2M} access
  technologies for mobile network operators,'' Tech. Rep., March 2014.

\bibitem{sched}
A.~Azari \emph{et~al.}, ``Lifetime-aware scheduling and power control for {M2M}
  communications in {LTE} networks,'' in \emph{IEEE VTC}, 2015, pp. 1--5.

\bibitem{nok}
{Nokia Networks}, ``Looking ahead to {5G}: Building a virtual zero latency
  gigabit experience,'' Tech. Rep., 2014.

\bibitem{cong}
F.~Cao and Z.~Fan, ``Cellular {M2M} network access congestion: Performance
  analysis and solutions,'' in \emph{IEEE Wireless and Mobile Computing,
  Networking and Communications conference}, Oct 2013, pp. 39--44.

\bibitem{laya}
A.~Laya, L.~Alonso, and J.~Alonso-Zarate, ``Is the random access channel of
  {LTE} and {LTE-A} suitable for {M2M} communications? a survey of
  alternatives.'' \emph{IEEE Commun. Surveys Tuts.}, vol.~16, no.~1, pp. 4--16,
  2014.

\bibitem{acb}
I.~Leyva-Mayorga \emph{et~al.}, ``Performance analysis of access class barring
  for handling massive {M2M} traffic in {LTE-A} networks,'' in \emph{IEEE ICC},
  May 2016, pp. 1--6.

\bibitem{ciot}
{3GPP TS 45.820}, ``Cellular system support for ultra-low complexity and low
  throughput internet of things (ciot),'' Tech. Rep., (Rel. 13).

\bibitem{TruAs}
R.~D. Gaudenzi \emph{et~al.}, ``Asynchronous contention resolution diversity
  {ALOHA}: Making {CRDSA} truly asynchronous,'' \emph{IEEE Trans. Wireless
  Commun.}, vol.~13, no.~11, pp. 6193--6206, Nov 2014.

\bibitem{twod}
Z.~Li \emph{et~al.}, ``{2D} time-frequency interference modelling using
  stochastic geometry for performance evaluation in low-power wide-area
  networks,'' \emph{arXiv preprint arXiv:1606.04791}, 2016.

\bibitem{ecra}
F.~Clazzer \emph{et~al.}, ``Exploiting combination techniques in random access
  {MAC} protocols: Enhanced contention resolution {ALOHA},'' \emph{arXiv
  preprint arXiv:1602.07636}, 2016.

\bibitem{ppz}
K.~Fyhn \emph{et~al.}, ``Multipacket reception of passive {UHF RFID} tags: A
  communication theoretic approach,'' \emph{IEEE Trans. Signal Process.},
  vol.~59, no.~9, pp. 4225--4237, 2011.

\bibitem{thes}
M.~T. Do, ``Ultra-narrowband wireless sensor networks modeling and
  optimization,'' Ph.D. dissertation, Lyon, INSA, 2015.

\bibitem{cfo}
J.~Fang \emph{et~al.}, ``Fine-grained channel access in wireless {LAN},''
  \emph{IEEE/ACM Trans. on Networking}, vol.~21, no.~3, pp. 772--787, 2013.

\bibitem{mrcegc}
Y.~Song \emph{et~al.}, ``Outage probability comparisons for diversity systems
  with cochannel interference in rayleigh fading,'' \emph{IEEE Trans. Wireless
  Commun.}, vol.~4, no.~4, pp. 1279--1284, 2005.

\bibitem{shif}
M.~Hua \emph{et~al.}, ``Analysis of the frequency offset effect on
  {Zadoff--Chu} sequence timing performance,'' \emph{IEEE Trans. Commun.},
  vol.~62, no.~11, pp. 4024--4039, 2014.

\bibitem{gb}
G.~Miao and G.~Song, \emph{Energy and Spectrum Efficient Wireless Network
  Design}.\hskip 1em plus 0.5em minus 0.4em\relax Cambridge University Press,
  2014.

\end{thebibliography}
\end{document}